\documentclass[]{spie}  
 
\newcommand{\R}{\mathbb{R}}
\newcommand{\z}{\mathbf{z}}
\newcommand{\f}{\mathbf{f}}

\usepackage{amsmath,amsfonts,amssymb}
\usepackage{graphicx}
\usepackage{subcaption}
\usepackage{placeins}
\usepackage{cite}
\usepackage{booktabs}
\usepackage[colorlinks=true, allcolors=blue]{hyperref}

\title{Evaluating Procedures for Establishing Generative Adversarial Network-based Stochastic Image Models in Medical Imaging}

\author[a]{Varun A. Kelkar}
\author[a]{Dimitrios S. Gotsis}
\author[b]{Frank J. Brooks}
\author[c]{Kyle J. Myers}
\author[c]{Prabhat KC}
\author[c]{Rongping Zeng}
\author[a,b]{Mark A. Anastasio}

\affil[a]{Department of Elect. \& Computer Eng., University of Illinois at Urbana-Champaign, IL, USA}
\affil[b]{Department of Bioengineering, University of Illinois at Urbana-Champaign, IL 61801, USA}
\affil[c]{Center for Devices and Radiological Health, Food and Drug Administration, Silver Spring, MD, USA}
\authorinfo{Send correspondence to Mark A. Anastasio (E-mail: maa@illinois.edu, Telephone: +1 (217) 300 0314) and Rongping Zeng (E-mail: Rongping.Zeng@fda.hhs.gov). }

\begin{document} 
\maketitle

\begin{abstract}
Modern generative models, such as generative adversarial networks (GANs), hold tremendous promise for several areas of medical imaging, such as unconditional medical image synthesis, image restoration, reconstruction and translation, and optimization of imaging systems. However, procedures for establishing stochastic image models (SIMs) using GANs remain generic and do not address specific issues relevant to medical imaging. In this work, canonical SIMs that simulate realistic vessels in angiography images are employed to evaluate procedures for establishing SIMs using GANs. The GAN-based SIM is compared to the canonical SIM based on its ability to reproduce those statistics that are meaningful to the particular medically realistic SIM considered. It is shown that evaluating GANs using classical metrics and medically relevant metrics may lead to different conclusions about the fidelity of the trained GANs. This work highlights the need for the development of objective metrics for evaluating GANs.
\end{abstract}

\keywords{Generative adversarial networks; objective image quality assessment; image perception; stochastic image models}

\section{Purpose}
\label{sec:purpose}  

Generative models, such as generative adversarial networks (GANs) are a class of models that seek to approximate unknown high-dimensional data distributions, for instance, image distributions \cite{gan_goodfellow, goodfellow_dl_book}. GANs hold promise for potential applications in medical imaging \cite{gan_mi_review}, such as unconditional medical image synthesis \cite{weimin_ambientgan, 3dstylegan}, image restoration and reconstruction \cite{inn_mri, gan_circle, picgm}, medical image translation \cite{medgan} and data augmentation \cite{gan_augmentation}. GANs have also been proposed as a tool for establishing stochastic image models (SIMs), with potential applications to objective assessment and optimization of medical imaging systems \cite{hmi, gam_mcmc, jha, evalsr}. Modern GANs such as the StyleGANs \cite{stylegan, stylegan2} are known to produce diverse, visually realistic images.

Despite the apparent realism and diversity in images, the extent to which a GAN learns image statistics is still an ongoing topic of research \cite{gan_evaluation, gan_test, gan_rucha}. In particular, image GANs may produce images that appear visually realistic, but possess ensemble statistics that are incorrect \cite{gan_rucha}, which could be detrimental to downstream applications \cite{hallucinations_sayantan}. The Frechet inception distance (FID) score, inception score (IS) and average log-likelihood are a few among a long list of metrics that have been proposed to evaluate GANs \cite{gan_evaluation}. However, these metrics are generic, as they do not address the specific issues relevant to medical imaging, and are not generally computed by use of those statistics that are of utility to an observer for performing a specified task. 
In this preliminary work, GANs are evaluated with the help of statistics that are relevant to medical imaging and are associated with a meaningful diagnostic task. For this purpose, canonical SIMs of realistic medical images are considered. GANs are trained on independent and identically distributed (IID) images from the canonical SIMs. The statistical quantities identified to be medically relevant are then computed on the images from the canonical SIM and the GAN-based SIM in order to evaluate the performance of the GAN.

\section{Methods}
\subsection{Canonical stochastic image models (SIMs) and relevant evaluation metrics}\label{sec:clb_mammographic}

In this work, the choice of the data distribution used to train the GAN was based on the following criteria. First, realistic canonical SIMs that are associated with a mathematical procedure for generating images were identified. This allows for direct control over image properties of interest. For these canonical SIMs, statistical quantities that are medically meaningful for the particular canonical SIM were identified. In this work, a canonical model of simulated angiograms having aneurysms at specified locations was used. A GAN was trained on images from the canonical SIM. A robust thickness estimation procedure measures the thickness of the vessel present in the angiograms at specified locations for both the canonical SIM and the GAN-approximated SIM. The estimated thickness of the vessel at the specified location, which is related to the presence of an aneurysm at the location, 
is chosen as the statistical quantity of interest that is used to evaluate the GAN.

The simulated angiogram SIM used in this study was inspired by the method developed by Rolland, \textit{et al.} \cite{rolland_angiogram}. Realistically shaped blood vessels were achieved by simulating a vessel trajectory that curves and twists in the three dimensional (3D) space as prescribed by the Frenet-Serret formulas \cite{frenet}. The curvature and torsion parameters of this space curve were set as specified in the Handbook of Medical Imaging (HMI) \cite{hmi}. For creating images with normal vessels without aneurysm, a thickness of 20 pixels was assigned throughout the length of the vessel, and accordingly, spheres of 20 pixels diameter were placed along the generated space curve. For creating images of vessels with aneurysm, the vessel thickness profile was set such that the thickness of the vessel at the specified aneurysm location is $t \sim \mathcal{N}(t_0, \sigma_{an})$ pixels, and the thickness tapers off to the normal vessel thickness of 20 pixels away from the aneurysm location, in accordance with the HMI \cite{hmi}. This was followed by a projection from the 3D space to a two-dimensional (2D) image. For simplicity, only one vessel was simulated for every image. In order to avoid images containing a very small portion of the space curve, it was ensured that the space curve always passes through the point that gets mapped to the center of the image after the projection. Additionally, the location of the aneurysm was also fixed to this point, and aneurysm was added to the vessel with a probability of 0.5. Other safeguards to preserve the realism of the images generated, such as clipping the rotation and torsion parameters to prevent excessive pigtailing, were added in accordance with the HMI \cite{hmi}. Edge noise was added in the form of random IID Gaussian noise in the vessel thickness along the vessel length. The value of $\sigma_{an}$ was chosen in accordance with the HMI. Three different SIMs were created such that for each of them, normal images and images with aneurysm occur with equal probability. Note that for each of the SIMs, the distribution of vessel thicknesses at the center of the image is a biomodal distribution, with the two modes centered at 20 pixels and $t_0$ respectively. The value of $t_0$ was 23, 27 and 33 pixels respectively for the three SIMs. These three SIMs represent different levels of aneurysm signal detectability in the angiogram. Measurement noise was not added to the simulated images.
Since the thickness of the vessel in an angiogram is a direct biomarker for the presence of an aneurysm, this study focuses on the distribution of vessel thicknesses as a medically relevant metric for evaluating the fidelity of the trained GAN.

\begin{figure}
     \includegraphics[width=\linewidth]{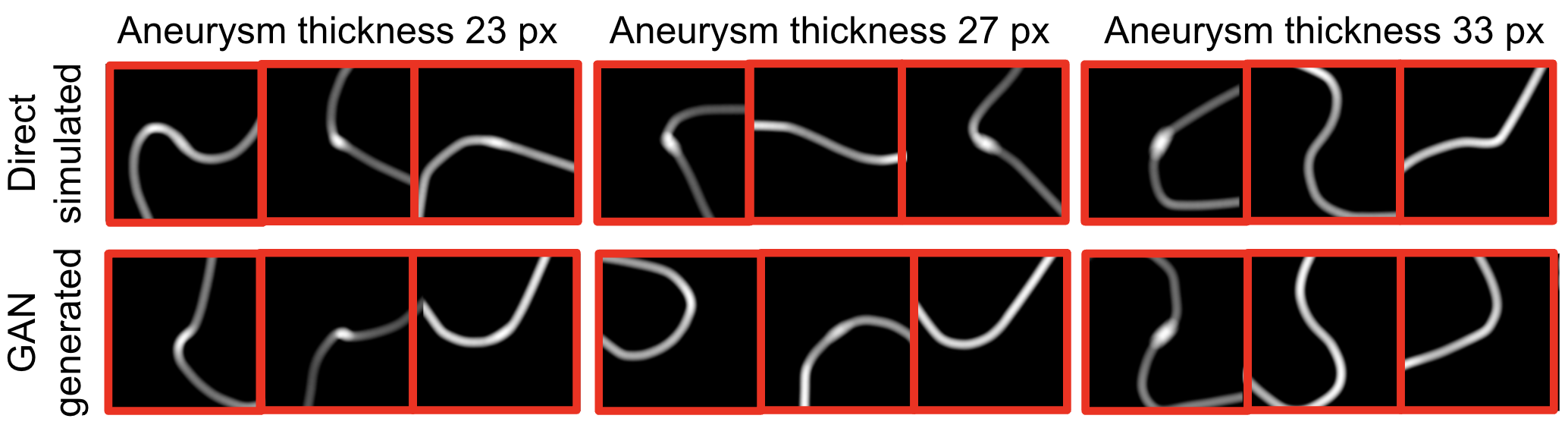}
      \caption{Direct simulated images from the angiogram SIMs and GAN-generated images}
      \label{fig:angiogram_real_and_fake}
\end{figure}

\begin{figure}
     \includegraphics[width=\linewidth]{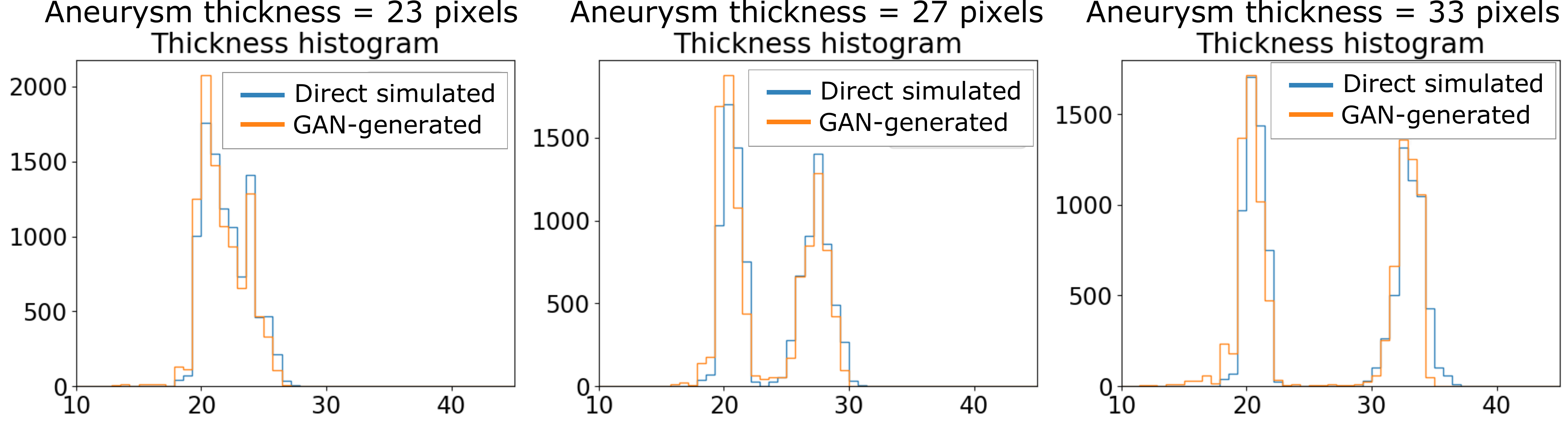}
      \caption{Empirical distributions of estimated thicknesses for the direct simulated and GAN-generated images for the three SIMs.}
      \label{fig:thickness_distributions}
\end{figure}

\subsection{Generative adversarial networks}
Generative adversarial networks (GANs) are a popular class of generative models that seek to approximate a data distribution by learning to map a sample $\z \in \R^k$ from a lower dimensional, tractable data distribution $p_{\z}$, such as the IID standard normal distribution, to a sample $\f$ from the high dimensional data distribution $p_{\f}$. 

In this work, one of the state-of-the-art GAN architectures, namely StyleGAN2, \cite{stylegan2} was employed to establish a learned SIM. Three separate StyleGAN2 models were trained, each on 100,000 images of size 256$\times$256 from one of the three canonical SIM models described above. A learning rate of 0.001 and a latent-space dimensionality of 512 was used for all of the models. Each of the models was trained using Tensorflow 1.14/Python 3.6 on two Nvidia Quadro RTX 8000 GPUs for 10000 iterations, with each iteration corresponding to 1000 images from the training dataset.

\subsection{Estimation of vessel thicknesses}\label{sec:features}
In this work, the images directly simulated from the canonical SIM will be termed as ``direct simulated" images.
The thickness of the vessels was estimated by use of a simple estimation technique described as follows. First, the angiogram image was thresholded and binarized. Next, the diameter of the largest circle that can be inscribed inside the vessel at the center of the image was computed. This was treated as the estimated thickness of the vessel at the central location. An empirical probability density function of the estimated per-image vessel thicknesses was established for both the canonical SIM and the GAN-approximated SIM. The resulting distributions were then used for further analysis in order to summarize trends. Two types of analyses were conducted -- the first computed an empirical Kullback–Leibler (KL) divergence between the thickness distributions computed from the direct simulated and GAN-generated images\cite{empirical_kl}. The second computed an empirical Jensen-Shannon (JS) divergence between the two thickness distributions. All the evaluation metrics were computed using 10000 direct simulated and GAN-generated images.

\section{Results and Discussion}
\begin{figure}[t]
\noindent\begin{minipage}{0.49\linewidth}
\captionsetup{type=figure}
\includegraphics[width=\textwidth]{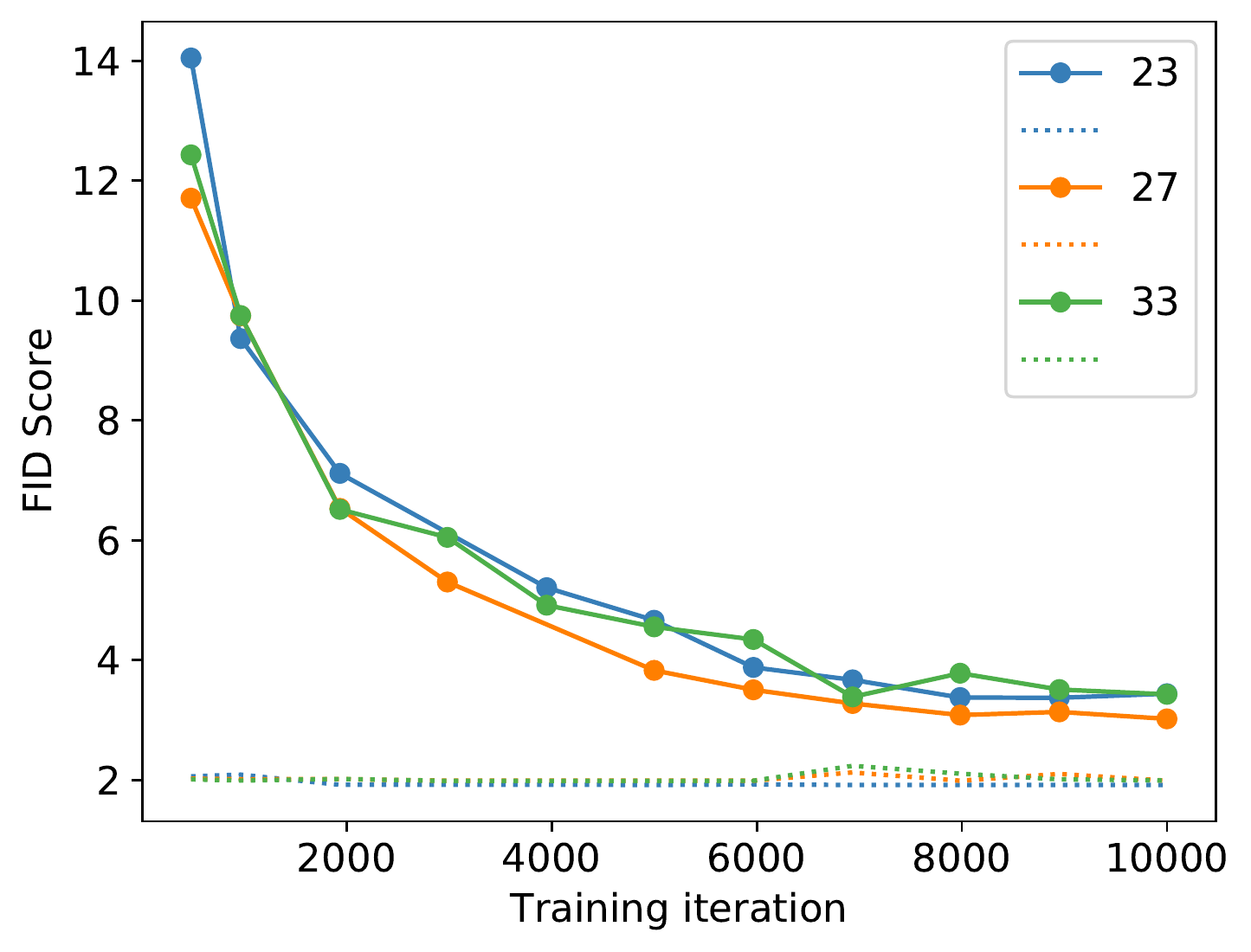}
\vspace{0.01cm}
\captionof{figure}{FID score between the direct simulated and the GAN-generated image distributions as a function of training iterations for SIMs with various aneurysm thicknesses. The dotted lines indicate the FID score computed using two independent direct simulated image datasets from the same distribution, and hence represents the variance in the FID estimator.}
\label{fig:fid_iters}
\end{minipage}\quad
\noindent\begin{minipage}{0.49\linewidth}
\captionsetup{type=figure}
\includegraphics[width=\textwidth]{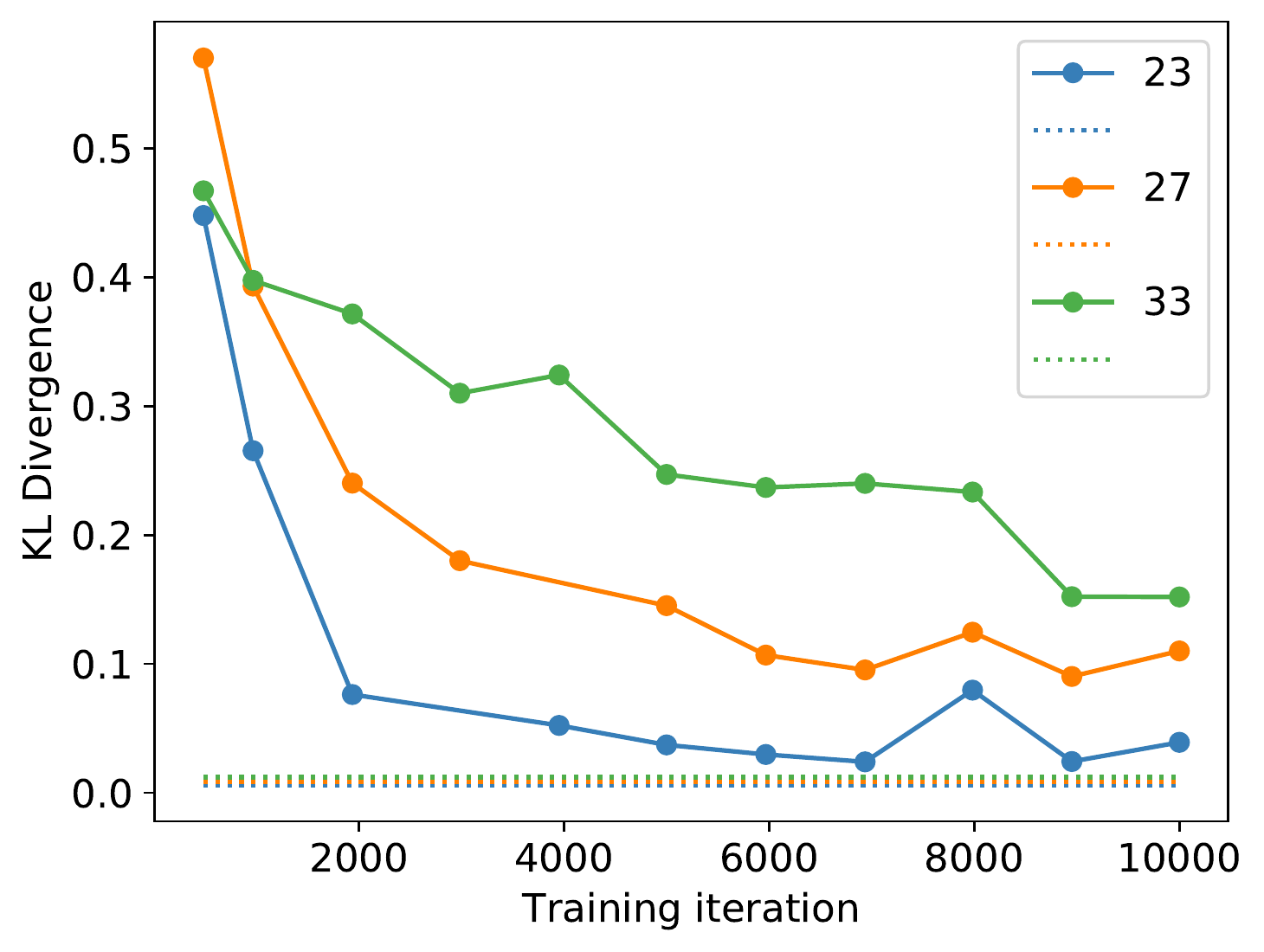}
\vspace{0.01cm}
\captionof{figure}{Empirical KL divergence between the thickness distributions computed from the direct simulated and GAN-generated images as a function of training iterations for SIMs with various aneurysm thicknesses. The dotted lines indicate the KL divergence computed using two IID thickness datasets from the direct simulated images, and hence represents the variance in the KL divergence estimator.}
\label{fig:kl_iters}
\end{minipage}
\end{figure}

\begin{figure}
\centering
\includegraphics[width=0.5\textwidth]{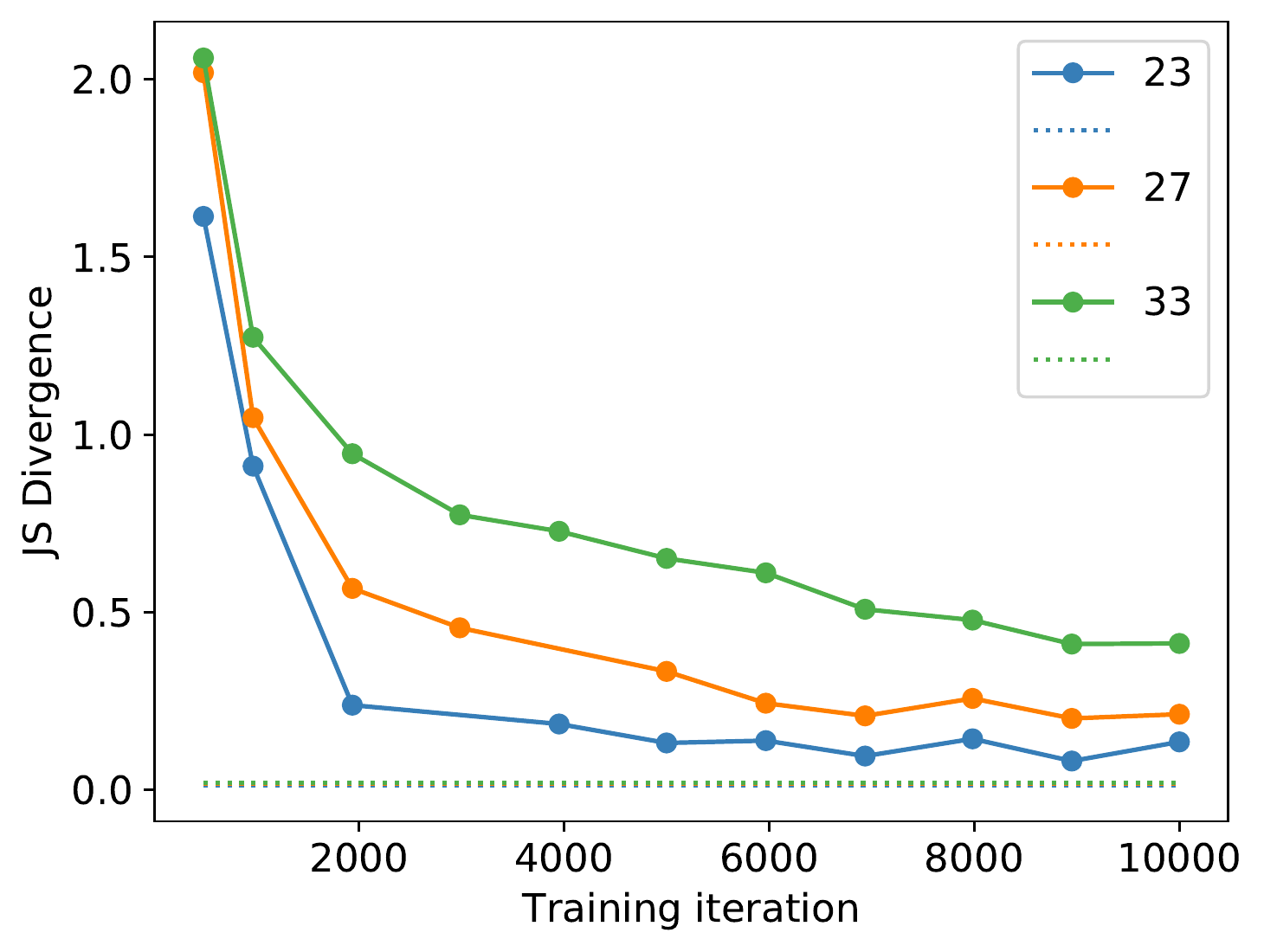}
      \caption{Empirical JS divergence between the thickness distributions computed from the direct simulated and GAN-generated images as a function of training iterations for SIMs with various aneurysm thicknesses. The dotted lines indicate the JS divergence computed using two IID thickness datasets from the direct simulated images, and hence represents the variance in the JS divergence estimator.}
      \label{fig:js_iters}
\end{figure}

Figure \ref{fig:angiogram_real_and_fake} shows the images generated by the GANs alongside the direct siimulated images from the training dataset. The FID score between images from the canonical SIMs and the learned SIMs was computed around every 1000 iterations, and is plotted in Fig. \ref{fig:fid_iters}. Figure \ref{fig:thickness_distributions} shows the empirical distribution of estimated thicknesses for the direct simulated and GAN-generated images after completing 10000 training iterations. It can be seen that, visually, the GAN seems to have learnt the class prevalence and the general shape of the distribution correctly. The match between the thickness distributions was also summarized quantitatively. Figure \ref{fig:kl_iters} shows the empirical KL divergence between the direct simulated and GAN-generated thickness distributions as a function of GAN training iteration. The dotted lines in Fig. \ref{fig:fid_iters}, Fig. \ref{fig:kl_iters} and Fig. \ref{fig:js_iters} respectively correspond to the FID score, empirical KL divergence and empirical JS divergence computed using two IID direct simulated datasets instead of a direct simulated and a GAN-generated dataset. Thus, this gives a measure of the noise floors in FID, KL and JS divergence estimations. The noise floors for the thickness KL and JS divergence score are very similar for the three SIMs. It can be seen that for all the three SIMs, 
the empirical KL and JS divergence approaches the noise floor, but still converges to a value higher than the noise floor. This phenomenon is increasingly more evident as the thickness of the aneurysm increases, indicating that out of the SIMs considered, the GAN trains better on the SIM with lower aneurysm thickness with respect to the KL and the JS divergence of the thickness distributions.
This, however, is in contrast to the findings from the FID scores shown in Fig. 2, which shows that the GAN trained on the three SIMs have approximately the same FID score. 

\vspace{5pt}

There are at least two reasons why the FID scores rank the three datasets in terms of the fidelity of trained GANs differently as compared to the metrics based on the identified thickness distributions. First, the FID score measures the Frechet distance in the feature space of an inception network \cite{inception} trained on the ImageNet dataset \cite{imagenet}. Hence, although commonly used, it is not tailored to the specific medical image distribution considered. Secondly, the GAN may learn the distribution of different features to different degrees of fidelity, resulting in different performance rankings when examined by different metrics. Both these hypotheses point to the need for motivating the choice of the computable statistical quantities themselves based on a task. While this requires significant effort to formulate, it also opens up the possibility of evaluating
GANs in terms of those statistics that influence task-performance, and makes setting up a formal, task-based evaluation pipeline easier. 


\vspace{-10pt}
\section{Conclusions}
In conclusion, this work presents preliminary results on the evaluation of procedures for establishing stochastic image models (SIMs) using GANs. The authors advocate the need to motivate the choice of the evaluation procedure based on statistical quantities that are meaningful for the particular task and the SIM considered.


\acknowledgments        
This research was supported in part by the FDA critical path funding. Varun A. Kelkar acknowledges funding by an appointment to the Research Participation Program at the Center for Device and Radiological Health administered by the Oak Ridge Institute for Science and Education through an inter-agency agreement between the U.S. Department of Energy and U.S. Food and Drug Administration.

\bibliography{report} 
\bibliographystyle{spiebib} 

\end{document}